\documentclass[aps,twocolumn,nofootinbib,prl,superscriptaddress]{revtex4}
\usepackage{psfrag}
\usepackage{amssymb}
\usepackage{graphicx}
\usepackage[mathscr]{eucal}

\newcommand{\vb}[1]{\mbox{{\boldmath$#1$}}\kern-.2em}

\newcommand{\tr}{{\rm tr}\,}

\newcommand{\ket}[1]{\vert #1 \rangle}
\newcommand{\bra}[1]{\langle #1 \vert}

 \newcommand{\eqref}[1]{(\ref{#1})}

\begin{document}
\title{Relative states, quantum axes and quantum references}
\author{E.~Bagan}
\altaffiliation[Permanent address: ] {Grup de F{\'\i}sica Te{\`o}rica \& IFAE, Fa\-cul\-tat de
Ci{\`e}ncies, Edifici Cn, Universitat Aut{\`o}noma de Barcelona, 08193
Bellaterra (Barcelona) Spain}
\affiliation{Physics Department, Brookhaven National Laboratory, Upton, New York 11973, USA}
\author{S.~Iblisdir}
\affiliation{GAP-Optique, Universit\'e de Gen\`eve, 1211 Gen\`eve, Switzerland}
\author{R.~Mu{\~n}oz-Tapia}
\affiliation{Grup de F{\'\i}sica Te{\`o}rica \& IFAE, Facultat de
Ci{\`e}ncies, Edifici Cn, Universitat Aut{\`o}noma de Barcelona, 08193
Bellaterra (Barcelona) Spain}

\begin{abstract}
We address the problem of measuring the relative angle between two `quantum axes' made out of~$N_1$ and~$N_2$ spins. Closed forms of our  fidelity-like figure of merit are obtained for an arbitrary number of parallel spins. The asymptotic regimes of large~$N_1$ and/or~$N_2$ are discussed in detail. The extension of the concept `quantum axis'  to more general situations is addressed. We give optimal strategies when the first quantum axis is made out of parallel spins whereas the second is a general state made out of two spins.
\end{abstract}

\maketitle

It has been traditionally assumed that data transmission through a quantum channel necessarily requires the existence of a reference shared by the sender (Alice) and the recipient (Bob).  Hence, if the rotational degrees of freedom of a system of spin-1/2 particles, for instance, are used as  quantum channel, it is commonly presumed that Alice and Bob share a spatial frame to which they refer their state components. 

Though such references can in principle be established via auxiliary quantum systems previous to the actual data transmission~\cite{frames}, the possibility of communication without them, using instead the orientation of some of the spins/qubits relative to others, has also been considered in the literature~\cite{comm1,comm2}. 
Only recently the advantages of encoding data in these relative degrees of freedom, such as robustness to certain kinds of noise, have started to emerge in this~\cite{comm2} and other contexts~\cite{cabello}, including computation~\cite{comp} and cryptography~\cite{cryp}.  As more attention is being paid to this topic, terms such as  `relative quantum information'  or simply, `relative states' have become part of the standard quantum information jargon.

This letter is concerned with estimation-theory aspects of relative states.  We wish to focus on the optimal estimation of the (relative) angle $\Theta$ formed by two `quantum axes' (QAs), by which we mean two systems of $N_i$ spin-1/2 particles ($i=1,2$ throughout the letter), each of them in a state $\ket{\Psi_i}$ possessing axial symmetry; i.e., invariant (up to a phase) under a rotation $R(\varphi_i \vec n_i)$ of arbitrary angle $\varphi_i$ about some unit (Bloch) vector $\vec n_i$. 

We will view one of these QAs (say, QA~\#1) as a `quantum reference'. In contradistinction to classical references, on which no (quantum) measurement  can be performed, Bob is allowed to perform joint measurements on both (reference) QA~\#1 and (signal) QA~\#2. We will pay particular attention to the detailed description of the transition from a quantum to a classical reference as the number of spins $N_1$ becomes very large.

This letter bears some similarities with previous work by Bartlett {\em et al.}~\cite{b+r+s}, where only product states (mostly parallel spins) for the two QAs are considered, and by Gisin {\em et al.}~\cite{g+i}, where, in addition, homodyne detection is also examined in the context of relative states. Here, the use of a fidelity-like figure of merit, instead of the information gain used in~\cite{b+r+s} or the mean variance of~\cite{g+i}, makes it possible to take our approach much further, without the need for numerical analysis.  Our analytical results reveal features that might have passed unnoticed (or might have been misidentified)  in these and other previous analysis~\cite{l+s+b}. Thus, e.g., we find that anti-parallel spins perform worse than parallel spins in all the situations considered in this letter. We also find that (and explain why) the optimal QAs are eigen-states of the total spin projection $\vec n_i\cdot\vec S_i$.
These features are generally believed to be independent of the figure of merit (or loss function) used in the analysis.

This letter is organized as follows. We first introduce the problem and our notation, paying special attention to the definition of a QA. We address the simple product-state case where the QAs consist of parallel spins. Next, the generalization to entangled states for QA~\#2  is considered. In particular the $N_2=1$ ($N_1$~arbitrary) case is solved analytically. We end up the letter with a brief summary and conclusions.

Let us start by assuming that Alice has sent $N_1+N_2$  ($N_i=2j_i$) spin-1/2 particles (spins for short) to Bob, with whom she does not share any reference frame. Let us further assume that the state of each of these  two sets of spins is given by $\ket{\Psi_i}=\ket{\psi_i}^{\otimes N_i}$ (parallel-spin case;  $N_i$ identical copies of the same state $\ket{\psi_i}$). Bob's task is to determine the overlap~$|\langle\psi_1|\psi_2\rangle|^2$ by performing generalized measurements on the state $\ket\Psi=\ket{\Psi_1}\otimes\ket{\Psi_2}$  of the $N_1+N_2$ spins. According to our discussion above, $|\langle\psi_1|\psi_2\rangle|^2$ is an example of relative quantum information.

One could look at this problem from a more geometrical point of view and say that Bob's task is to estimate the angle~$\Theta$ defined as
\begin{equation}
\Theta=\arccos\left[2|\langle\psi_1|\psi_2\rangle|^2-1\right] ,
\label{e-def Theta}
\end{equation}
which is twice the Fubini-Study distance between~$\ket{\psi_1}$ and~$\ket{\psi_2}$~\cite{fub},  and also coincides with the angle between their corresponding Bloch vectors~$\vec n_1$ and~$\vec n_2$. From this point of view, the states $\ket{\Psi_i}$ define two QAs, and Bob's task is to estimate the angle $\Theta$ between them.

Like their classical counterparts, $\ket{\Psi_i}$ possess a manifest axial symmetry i.e., they are invariant under the $SU(2)$ transformations $U(\varphi_i \vec n_i)=[u(\varphi_i \vec n_i)]^{\otimes N_i}$ [and so is $\Theta$, as can be readily seen from~(\ref{e-def Theta})],  which are representations of the rotations~$R(\varphi_i \vec n_i)$ in the Hilbert spaces~${\mathscr H}_i^{\otimes N_i}$, where ${\mathscr H}_i={\mathbb C}^2$.
This (axial) symmetry provides a means of extending the notion of QA, and thus of the angle between two such QAs. We will come back to this point in the second part of this letter.

To quantify the quality of Bob's estimation procedure, we use the average of the following figure of merit
\begin{equation}
\Delta(\Theta,\Theta_\chi)\equiv \cos(\Theta-\Theta_\chi) ,
\label{e-fom}
\end{equation}
where~$\Theta_\chi$ is Bob's estimate of~$\Theta$, based on a particular outcome~$\chi$ of his measurement, represented by a positive operator valued measure (POVM)~$\{\tilde O_\chi\}$. From~$\Delta(\Theta,\Theta_\chi)$ one can readily compute the fidelity $F(\Theta,\Theta_\chi)=[1+\Delta(\Theta,\Theta_\chi)]/2$, which is the figure of merit used in~\cite{l+s+b}.
The average is defined as
\begin{equation}
\bar\Delta=\sum_\chi\int dg_1\,dg_2\, \Delta[\Theta(g_1,g_2),\Theta_\chi] \tr[\tilde O_\chi\rho(g_1,g_2)] ,
\label{e-F}
\end{equation}
where  in writing~(\ref{e-F}) the following arguments have been taken into account. First, relative to an observer's reference frame (of which neither Alice nor Bob need be aware), we have
\begin{equation} 
\ket{\Psi_i}=U(g_i)\ket{\Psi^{(0)}_i}   
\label{e-Psi 0}
\end{equation}
for two suitable triplets of Euler angles $g_i=(\alpha_i,\beta_i,\gamma_i)$,
where $\vec z\cdot\vec S\,\ket{\Psi^{(0)}_i}=j_i \ket{\Psi^{(0)}_i}$ (i.e., $ \ket{\Psi^{(0)}_i}=\ket{j_ij_i}$ in the standard notation). Note that  $\rho(g_1,g_2)\equiv \ket{\Psi}\bra{\Psi}$ is actually independent of $\alpha_1$ and $\alpha_2$. Second, since Bob has no prior knowledge about $\ket{\psi_i}$, the 
{\em a priori} probability  for each of these states is
$dg_i$, which stands for the invariant Haar measure of~$SU(2)$ [for these states, which are independent of $\alpha_i$,  
one can use the normalized solid angle $d\Omega_i/(4\pi)$ instead of $dg_i$]. Third, $\Theta$ is now denoted $\Theta(g_1,g_2)$. Actually,  $\Theta(g_1,g_2)=\Theta(e,g_1^{-1}g_2)$,  where $e$ stands for the null Euler angles, i.e., $U(e)=\openone$ (the remaining notation is conventional). This shows that $\Theta$ depends only on the relative parameter $g_1^{-1}g_2$, as it should be.

Using rotational covariance, one can integrate out the global parameters $g'=g_1$ and express $\bar\Delta$ just in terms of the relative ones $g=g_1^{-1}g_2$~\cite{b+r+s} as
\begin{equation}
\bar\Delta=\sum_\chi\int dg\, \Delta[\Theta(e,g),\Theta_\chi] \tr[O_\chi\rho(e,g)] ,
\label{e-F rel}
\end{equation}
where $O_\chi$ is the result of rotating $\tilde O_\chi$ into all possible global orientations (labeled by~$g'$) and, as such, it is rotationaly invariant. Shur's lemma and elementary convexity arguments further tell us that $\{O_\chi\}$ can be chosen to be the projectors $\{\openone_j\}$ on the spin-$j$ subspaces ${\mathscr H}^{(j)}$, $|j_1-j_2|\le j\le j_1+j_2$,  where $SU(2)$ acts irreducibly~\cite{b+r+s}.

The advantage of $\Delta(\Theta,\Theta_\chi)$ over other figures of merit becomes apparent when we re-cast it as the scalar product of two 2-dim real vectors, namely, as $\Delta[\Theta(e,g),\Theta_\chi]=\vb w\cdot\vb w_\chi$, where $\vb w=(\cos\beta,\sin\beta)$, $\vb w_\chi=(\cos\Theta_\chi,\sin\Theta_\chi)$, and we have used that $\Theta(e,g)=\beta$. By recalling the Schwarz inequality, one obtains that the optimal estimate is $\vb w_\chi=\vb V_\chi/|\vb V_\chi|$, where 
\begin{equation}
\vb V_j=\int dg  \, \vb w  \;\tr[\openone_j \rho(e,g)]
\end{equation}
(recall that $\chi=j$ in the parallel spin case under discussion),
and the maximum value of $\bar\Delta$  is
\begin{equation}
\bar\Delta=\sum_j |\vb V_j|.
\label{e-Delta}
\end{equation}

The probabilities $\tr[\openone_j \rho(e,g)]$ can be straightforwardly computed in terms of Clebsch-Gordan coefficients, $\langle jmj'm'|JM\rangle$, and the reduced Wigner matrices ${\rm d}^{(j)}_{mm'}(\beta)$~\cite{edm}. One obtains
\begin{equation}
\kern-.6em\tr [ \openone_j \rho(e,g)]\!=\!\sum_m\!\left[\langle j_1j_1j_2m-j_1|jm\rangle
{\rm d}^{(j_2)}_{m-j_1\,j_2}(\beta)\right]^2  ,
\end{equation}
where the trivial integration over the Euler angles~$\alpha$ and~$\gamma$ has already been carried out,  and here and below the sums extend over those values of the  dummy indexes  for which the expressions makes sense.
Integrating now~$\beta$ we obtain
 \begin{equation}
\vb V_j=\sum_m {\langle j_1j_1j_2m-j_1|jm\rangle^2\over(2j_2+1)(j_2+1)}\pmatrix{c^{j_2}_{m-j_1} \cr
 s^{j_2}_{m-j_1}}  ,
 \end{equation}
 where
\begin{equation}
c^{j_2}_m=m;\quad
s^{j_2}_m
={\Gamma({3\over2}+j_2+m)\Gamma({3\over2}+j_2-m)\over\Gamma(1+j_2+m)\Gamma(1+j_2-m)}.
\label{e-sj2m w gamm}
\end{equation}
For  later convenience we define  $\eta^{j_2}_{jm}= \langle j_1j_1j_2m-j_1|jm\rangle^2$ (the $j_1$ dependence is understood), which can be expressed as the ratio of factorials 
 \begin{eqnarray}
&&\kern-2.em \eta^{j_2}_{jm}
= {(2j+1)(2j_1)!(j_2-j_1+j)!\over(j_1-j_2+j)! (j_1+j_2-j)!(j_1+j_2+j+1)!}\nonumber\\[.5em]
&&\phantom{\kern-2.em \eta^{j_2}_{jm}}\times
{(j_2+j_1-m)!(j+m)!\over (j_2-j_1+m)!(j-m)!}  .
 \end{eqnarray}
 With this, $\bar\Delta$ can be written as
 \begin{eqnarray}
&&\kern-2.5em\bar\Delta={1\over(2j_2+1)(j_2+1)} \nonumber\\[.5em]
 &&\kern-2.5em\times\kern-.5em
 \sum_{l=-j_2}^{j_2}\!\!\sqrt{\left[\sum_k\eta^{j_2}_{l+j_1\,k+j_1} \, s^{j_2}_{k}\right]^2\!\!+\left[\sum_k\eta^{j_2}_{l+j_1\, k+j_1}\, c^{j_2}_{k}\right]^2}   ,
 \label{e-Delta par}
 \end{eqnarray}
 where we have assumed without any loss of generality that~$j_2\le j_1$. 
 
 As $N_1$ ($j_1$) becomes large (the reference QA~\#1 `becomes classical') we see that
 \begin{equation}
 \eta^{j_2}_{l+j_1\,k+j_1}\sim (2j_1)^{k-l}  .
 \label{e-eta asymp}
 \end{equation}
Hence, $ \eta^{j_2}_{l+j_1\,k+j_1}$ vanishes  at a rate $j_1^{-1}$ or faster [$k\le l$ in~Eq.~(\ref{e-Delta par})] unless $k=l$, in which case~$ \eta^{j_2}_{l+j_1\,l+j_1}\to1$.  We have
\begin{equation}
\bar\Delta=\bar\Delta_{\infty}-{\kappa_{j_2}\over N_1}+o(N_1^{-1})  
\label{e-Delta asymp}
\end{equation}
(recall that $N_i=2j_i$), where
\begin{equation}
\bar\Delta_{\infty}={1\over(2j_2+1)(j_2+1)}\sum_{m=-j_2}^{j_2}\sqrt{(s^{j_2}_m)^2+( c^{j_2}_m)^2} ,
\label{e-class short}
\end{equation}
and the coefficient~$\kappa_{j_2}$ are given by
\begin{equation}
\kappa_{j_2}={1\over(2j_2+1)(j_2+1)}\sum_{l=-j_2}^{j_2}{A^{j_2}_l c^{j_2}_l-B^{j_2}_l (s^{j_2}_l)^2\over\sqrt{(s^{j_2}_l)^2+( c^{j_2}_l)^2} },
\end{equation}
where 
\begin{eqnarray}
A^{j_2}_l&=&j_2(j_2+1)-l(3l-1);\\
B^{j_2}_l&=&{6l(j_2+l)+l-j_2\over 2j_2+2l+1} .
\end{eqnarray}

Eq.~(\ref{e-class short})
gives  the maximum value of our averaged figure of merit that can be attained  with an unlimited number of parallel spins  as a reference QA. Is this entirely equivalent to replacing QA~\#1 by a classical axis (a classical system)? This question deserves a few words. If the angle $\Theta$ is referred to a classical axis and Alice makes it  available to Bob, obviously the quality $\bar\Delta_{\rm class}$ of his estimation procedure  cannot be less than $\bar\Delta_\infty$. 
This is so because Bob could prepare an arbitrary number  of ancilla spins in the state $\ket{j_1j_1}$ and make them play the role of QA~\#1. He could then perform on the whole system, consisting of the ancill\ae\ and QA~\#2, the optimal measurement given by~$\{\openone_j\}$, which can be thought of a generalized measurement on QA~\#2 alone. 

But Bob could prepare his ancill\ae\ in a much more general state (such as a generalization of those in the second part of this letter). Could he not attain a larger $\bar\Delta_{\rm class}$ by proceeding this way?
To answer this question we go back to~(\ref{e-F rel}) and replace the trace by
\begin{equation}
\bra{\Psi_2}O_\chi\ket{\Psi_2}=\sum_m [O_\chi]_{mm} \left[
{\rm d}^{(j_2)}_{m\,j2}(\beta)\right]^2   ,
\label{e-prob class}
\end{equation}
as corresponds to replacing the QA~\#1 by the (classical) $\vec z$~axis,
where $[O_\chi]_{mm'}=\bra{j_2m}O_\chi\ket{j_2m'}$ and the integration over~$\alpha$ and~$\gamma$ has been brought forward.
We obtain
 \begin{equation}
 \kern-.5em
\vb V_\chi=\sum_m { [O_\chi]_{mm}  \over(2j_2+1)(j_2+1)}\pmatrix{c^{j_2}_{m} \cr
 s^{j_2}_{m}} \!\! \equiv \!\! \sum_m [O_\chi]_{mm}  \vb V_{m} ,
 \end{equation}
 and have the bound
 \begin{equation}
 \bar\Delta_{\rm class}=\sum_\chi|\vb V_\chi|\le \sum_m \left(\sum_\chi [O_\chi]_{mm} \right) |\vb V_m| .
 \end{equation}
Since $O_\chi$ is a POVM on ${\mathscr H}_2^{\otimes N_2}$, the sum in parenthesis is unity for all $m$, and the remaining sum becomes~(\ref{e-class short}). We conclude that $\bar\Delta_{\rm class}=\bar\Delta_{\infty}$. Moreover, this is attained with the von Neumann's measurement $O_m=\ket{j_2m}\bra{j_2m}$, i.e., by measuring the total spin along the classical axis with a Stern-Gerlach.

One could also consider the asymptotic regime when $N_2$ ($j_2$) is large but still $N_2\ll N_1$. A rather tedious calculation
yields
\begin{equation}
\bar\Delta_{\rm class}=\bar\Delta_{\infty}=1-{1\over2N_2}+o(N_2^{-1}), 
\label{e-Delta class asymp}
\end{equation}
as could be expected on statistical grounds~\cite{stat}. The limit~$j_2\to\infty$ of the sub-leading term in~(\ref{e-Delta asymp}) can be computed in a similar fashion to obtain the simple result~$\kappa_\infty=1/2$.

So far, we have just considered product states for both the reference and the signal QAs. We now wish to generalize the notion of QA to include entangled states, which are known to perform better as direction indicators~\cite{dir ind}. One should, likewise, expect them to provide a significant  improvement in the problem at hand. 

As discussed in the introductory part of this letter, the states $\ket{\Psi_i}$ are invariant under $U(\varphi_i\vec n_i)$ (they possess axial symmetry). This can be used to define a general~QA
in the following way: {\em a QA is
a state $\ket{\Psi_n}\in({\mathbb C}^2)^{\otimes N}$ satisfying
\begin{equation}
{}\left[ U(\varphi \vec n),\ket{\Psi_n}\bra{\Psi_n}\right]=0
\end{equation}
for some unit vector $\vec n$ and {\em any} value of~$\varphi$. }

Recalling that $U(\varphi\vec n)=\exp(-i\varphi\,\vec n\cdot\vec S)$, where $\vec S$ is the total spin operator, one finds that all
QAs are necessarily  of the form
\begin{equation}
\ket{\Psi^M_n}=\sum_{J=|M|}^{N/2} a_J\ket{JM}_n=U(g)\, \sum_{J=|M|}^{N/2}  a_J\ket{JM} 
\label{e-Psi M n}
\end{equation}
(no sum over the magnetic quantum number $M$), where~$g$ denotes the Euler angles of the rotation that takes~$\vec z$ to~$\vec n$ and the rest of the notation should be obvious. In other words, QAs are eigen-states of $\vec n\cdot\vec S$ with eigen-value~$M$. This should not come as a surprise, since optimal direction indicators are also of this kind~\cite{dir ind}.

For two QAs, $\ket{\Psi_i}\equiv\ket{\Psi_{n_i}^M}$ 
one can generalize~(\ref{e-def Theta}) as
\begin{equation}
\Theta=\arccos \left(\vec n_1\cdot \vec n_2 \right)  .
\end{equation}
Thus, it still makes perfect sense to talk about their relative angle (or Fubini-Study distance), even though~(\ref{e-def Theta}) does not apply. 

A final remark about Eq.~(\ref{e-Psi M n}). In the Clebsch-Gordan decomposition of~$({\mathbb C}^2)^{\otimes N}$, implicit in~(\ref{e-Psi M n}), the multiplicity  $n_J$ of each spin-$J$ representation has been
ignored. This does not imply any loss of generality. One can check that the orthogonal complement of the set of states $\{U\sum_\lambda b_\lambda \ket{JM\lambda}\}_{U\in SU(2)}$, where $\lambda$ labels the equivalent spin-$J$ representations, has dimension $(2J+1)\times(n_J-1)$, thus showing that this set spans only {\em one} spin-$J$ invariant subspace. 

Now we can consider a situation where QA~\#1 still consists of $N_1$ parallel spins but QA~\#2 has been generalized according to~(\ref{e-Psi M n}). One can still write~$\ket{\Psi_i}$ as in~(\ref{e-Psi 0}), where
$\vec z\cdot \vec S\ket{\Psi^{(0)}_1}=j_1\ket{\Psi^{(0)}_1}$ and
$\vec z\cdot \vec S\ket{\Psi^{(0)}_2}=M\ket{\Psi^{(0)}_2}$. Hence, the state $\rho(g_1,g_2)$ of the system made out of the two QAs is still manifestly independent of $\alpha_1$ and $\alpha_2$. Eq.~(\ref{e-F rel}) also holds, but the rotational invariance of the POVM now implies a more complex structure for the operators~$O_\chi$. They can be
chosen as 
\begin{equation}
O_{r j}=\sum_{JJ'}\xi^{J'J}_{r j} \openone^{j_1\otimes J'}\openone_j\openone^{j_1\otimes J}    ,
\end{equation}
where $\openone^{j_1\otimes J} $ is the identity in ${\mathscr H}^{(j_1)}\otimes{\mathscr H}^{(J)}$ and the pair $(r,j)$ plays the role of~$\chi$.
The coefficients~$\xi^{J'J}_{r j}$ satisfy
\begin{equation}
\kern-0.em\sum_r \xi_{r j}^{J'J}=\delta^{J'J} \times
\left\{
\begin{array}{ll}
1&\mbox{if} \  |j_1-J|\le j\le j_1+J;\\
0&\mbox{otherwise;}
\end{array}
\right.
\label{e-Sum r xi}
\end{equation}
and one can check that $\sum_{rj} O_{rj}=\openone$. 
For fixed $(r,j)$, the coefficients $\xi^{J'J}_{rj}$ can be viewed as a matrix  
with rows and columns labeled by $J'$ and $J$ respectively.
The positivity of the POVM operators $O_{rj}$ is equivalent to the positivity~of this matrix.
Our figure of merit is given by
\begin{equation}
\bar\Delta^M=\sum_{l=-j_2}^{j_2}  \sum_r  |\vb V^{\,M}_{r \, j_1+l}|  ,
\label{e- Delta sum mag}
\end{equation}
where
\begin{equation}
\vb V_{r j}^{\,M}=\sum_{JJ' k} a_{J'}^*a_J\,\xi^{J'J}_{r j} \sqrt{\eta^{J'}_{j\,k+j_1}\eta^{J}_{j\,k+j_1}} 
\pmatrix{C^{J'J}_{kM}
\cr
S^{J'J}_{kM} 
}  ,
\label{e-V r j}
\end{equation} 
and  we have used the definition
\begin{equation}
\pmatrix{C^{J'J}_{kM}\cr
S^{J'J}_{kM}}
=
\int_0^\pi{d\cos\beta\over2}\, \vb w \;\, {\rm d}^{(J')}_{kM}(\beta){\rm d}^{(J)}_{kM}(\beta)
.
\label{e-def C S}
\end{equation}
The coefficients $c^{j_2}_k$ and $s^{j_2}_k$ are special cases of~(\ref{e-def C S}): 
$C^{JJ}_{kJ}=(2J+1)(J+1)c^{J}_k$ and  $S^{JJ}_{kJ}=(2J+1)(J+1)s^{J}_k$. Closed expressions for $C^{JJ'}_{kM}$ can be given in terms of Clebsch-Gordan coefficients~\cite{edm} but, as far as we are aware, no such closed expressions exist for~$S^{JJ'}_{kM}$. Note that $\bar\Delta$ in~(\ref{e-Delta par}) coincides with $\bar\Delta^{j2}$.

The asymptotic expression of $\bar\Delta^M$ when $N_1$ becomes large, i.e., $\bar\Delta^M_\infty$, can be obtained straightforwardly using~(\ref{e-eta asymp}).  As for the parallel-spin case, one can workout~$\bar\Delta^M_{\rm class}$ by generalizing~(\ref{e-prob class'}), which now reads
\begin{equation}
\bra{\Psi_2}O_\chi\ket{\Psi_2}=\sum_{JJ'm} [O_\chi]^{JJ'}_{mm} \,
{\rm d}^{(J)}_{mM}(\beta)
{\rm d}^{(J')}_{mM}(\beta) .
\label{e-prob class'}
\end{equation}
One realizes that the optimal POVM is of the form $O_{r m}$, i.e., $\chi$ must necessarily include the magnetic number $m$, and
after a straightforward calculation one finds that $\bar\Delta^M_\infty=\bar\Delta^M_{\rm class}$. Hence, a reference QA made out of parallel spins is asymptotically optimal. For finite $N_1$ there might be better references than $\ket{j_1j_1}$, but as $N_1$ becomes larger, the additional quality they might provide gradually disappears.

This is just about as far as one can get for arbitrary $j_1$ and $j_2$. The rest of the letter deals with the particular case $N_2=2$  ($N_1$ arbitrary), which is simple enough to be solved analytically, but yet reveals some important features of the problem. The case $M=\pm1$ corresponds to the parallel-spin analysis carried out in the first part of this letter. We here focus on~$M=0$.

Since $j_2=1$, the index $l$ in~(\ref{e- Delta sum mag}) can take three values, namely $l=0,\pm1$, whereas $J$ and $J'$ can only take two: $J,J'=0,1$. For simplicity we write $a_1^2=x$ and $a_0^2=1-x$, where $0\le x\le 1$ to ensure the correct normalization of $\ket{\Psi_2}$, and we have used that $a_J$ can be chosen to be positive, since the phases can be absorbed in the POVM elements $\xi^{J'J}_{rj}$ [see~(\ref{e-V r j})]. 
With these definitions, and recalling~(\ref{e-Sum r xi}), we have
\begin{equation}
\sum_{l=\pm1}\sum_r|\vb V^{\;0}_{r \,l+j_1}|=\left(\sum_r \xi^{11}_{r\, l+j_1} \right) \pi{ax\over4}=\pi{ax\over4} ,
\label{e-V chi j1 pm 1}
\end{equation}
and
\begin{eqnarray}
\sum_r|\vb V^{\; 0}_{r j_1}|\!\!&=&\!\!
\sum_r \left\{
bx (1-x)\left[{\Re\,{\xi^{01}_{\chi j_1}}}\right]^2\phantom{\pi^2\over4^2} \right. \nonumber\\
&+&\!\!\!\!\left. {\pi^2\over4^2}\left[(1-x)\xi^{00}_{r j_1}\!+(1\!-a)x \,\xi^{11}_{r j_1}\right]^2 \right\}^{1\over2}
\kern-.5em ,
\label{e-V chi j1}
\end{eqnarray}
where
\begin{equation}
a={6j_1+5\over8(j_1+1)};\quad b={4j_1\over9(j_1+1)} .
\end{equation}
We can now use the constrain
$(\Re\,\xi^{01}_{\chi j_1})^2\le|\xi^{01}_{\chi j_1}|^2\le \xi^{00}_{\chi j_1}\xi^{11}_{\chi j_1} $,
where the last inequality follows from the positivity of $O_{r j}$, 
to obtain the bound
\begin{equation}
\sum_r|\vb V^{\; 0}_{r j_1}|\le 
 \sqrt{
b x (1-x)
+{\pi^2\over4^2}(1-ax)^2 }.
\label{e-V chi j1 bound}
\end{equation}
This bound is attainable, as one can check  by substituting in~(\ref{e-V chi j1}) the simple instance $\xi^{00}_{r j_1}=\xi^{11}_{r j_1}={1/2}$,  $\xi^{01}_{r j_1}=\xi^{10}_{r j_1}={(-1)^r/2}$;  $r=1,2$.
Adding~(\ref{e-V chi j1 pm 1}) and~(\ref{e-V chi j1}) we find
\begin{equation}
\bar\Delta^0(x)={\pi\over4}a x+\sqrt{b x(1-x)+{\pi^2\over4^2}(1-a x)^2} .
\label{e-Delta general}
\end{equation}
{}From this expression one can easily work out the maximum value of $\bar\Delta^0$,  which is
\begin{equation}
\bar\Delta^0_{\rm max}={a\pi(8b-a\pi^2)+16b\sqrt{4b+(1-a)\pi^2}\over 4(16b-a^2\pi^2)}.
\label{e-Delta general max}
\end{equation}

These results, in particular Eq.~(\ref{e-Delta general max}), can be now compared with Eq.~(\ref{e-Delta par}) (for $j_2=1$) to reach the following conclusions:  (i)~$\bar\Delta_{\rm par}\equiv\bar\Delta^{\pm1}\equiv\bar\Delta_{\rm Eq.(\protect\ref{e-Delta par})} < \bar\Delta^0_{\rm max}$, for any value of $j_1>0$, i.e., QAs with~$M=0$ are better `relative-direction indicators' than those with~$M=\pm1$ (parallel spins); (ii)~$\bar\Delta_{\rm anti}\equiv\bar\Delta^0(x=1/2) < \bar\Delta_{\rm par}$, i.e., anti-parallel spins (for which $x=1/2$) provide a less accurate estimation of~$\Theta$ than parallel spins. This is surprising, because two spins are known to encode more information when they are anti-parallel~\cite{gisin-popescu}.

For $N_1=1$ one finds 
\begin{equation}
\bar\Delta_{\rm anti}={\pi\over12}+{1\over6}\sqrt{{4\over3}+\pi^2} =0.81965. 
\end{equation}
Note that this figure is remarkably close to
 \begin{equation}
\bar\Delta_{\rm par}={2 \sqrt{4+9\pi^2}   +    \sqrt{16+9\pi^2}     \over36} =0.81966,
 \end{equation}
 which explains why numerical analysis fails to reveal any difference between parallel and anti-parallel spin~QAs~\cite{g+i,l+s+b}.

The classical limit, $\bar\Delta^0_\infty=\bar\Delta^0_{\rm class}$, is readily  seen to be given by $a=3/4$, $b=4/9$. Features (i) and (ii)  persist in this asymptotic regime.

At this point, one may wonder whether the notions of~QA  and the angle between two of them can be further generalized. To convince ourselves that this is not possible, let the states $\ket{\Psi_i^{(0)}}$ in~(\ref{e-Psi 0}) be completely general combinations of~$\ket{JM}$. Regardless the precise way we associate a relative angle~$\Theta$ to these two states, we must associate the {\em very same} value to~$U(\alpha,0,0)\ket{\Psi_i^{(0)}}$, i.e., to any state obtained by rotating~$\ket{\Psi_i^{(0)}}$ and arbitrary (Euler) angle~$\alpha$ about the $z$-axis of the observer's reference frame. Failure to do so, will clearly result in an ill-defined~$\Theta$, which will depend on the particular choice of 
this frame (more precisely, on the choice of axes $x$ and~$y$).
The contribution of all these states to ~$\bar\Delta$  in~(\ref{e-F rel}) becomes that of the mixed state
\begin{eqnarray}
\rho^{(0)}&=&\int_0^{2\pi}{d\alpha\over2\pi} U(\alpha,0,0) \ket{\Psi_2^{0}}\bra{\Psi_2^{0}} U^\dagger(\alpha,0,0)
\nonumber\\
&=&\sum_M p_M \ket{\Psi_z^M}\bra{\Psi_z^M}  ,
\label{e-convex set}
\end{eqnarray}
where the second equality follows from Schur's lemma, $0\le p_M\le1$ (it is a probability), and $\ket{\Psi_z^M}$ are QA with magnetic number $M$ 
relative to the observer's $z$ axis; these are precisely the type of states $\ket{\Psi_i^{(0)}}$ used in the definition of~$\bar\Delta^M$.
Because of the general structure $\bar\Delta=\sum_\chi|\vb V_\chi|$ of our figure of merit for the optimal estimator,  Eq.~(\ref{e-convex set}) translates into $\bar\Delta\le \sum_M p_M \bar\Delta^M\le \max_M\{\bar\Delta^M\}$, and we see that 
states with well defined magnetic number are optimal.
According to very general convexity arguments, this is also the case for any other convex figure of merit, such as the fidelity or the information gain. We conclude that~(\ref{e-Psi M n}) is the only sensible generalization of the notion of~QA.

In summary, we have analyzed the problem of estimating the angle between two directions (axes) in the absence of a shared reference frame and when these directions are represented by spin systems (quantum axes). In the case where the two directions are specified  by a number of parallel spins ($N_1$ and $N_2$ respectively), closed expressions for our fidelity have been given, and we have studied in detail the asymptotic regime of large~$N_1$ and/or large~$N_2$. One of our results is that when~$N_1$ is very large and $N_2\ll N_1$, optimal measurements tend to a Stern-Gerlach measurement on the second set of spins along the axis specified by the first set. We have also analyzed the situation where the first direction is given by~$N_1$ identically prepared spin system, and the second system is in a general eigenstate of the (projected) total spin [see Eq.(\ref{e-Psi M n})]. This analysis has allowed us to shed new light on the comparison between the performance of parallel vs. anti-parallel spin pairs. Finally, we have argued why states which are not of the form (\ref{e-Psi M n}) are irrelevant to relative state estimation. 

To conclude, let us mention a few open questions raised by our work. We have here analyzed the performance of a quantum system as a single reference \emph{axis}.  
One could try to extend our work and use such systems as complete reference  \emph{frames}, relative to which the full state of another 
quantum system could be estimated. 
Another issue which has not been addressed by our work is noise. The robustness of relative information hints at the convenience of using quantum references in noisy channels. 
Extensions of our work can help identify robust reference axes/frames, which is a compelling task. 

We acknowledge useful discussions with N.~Gisin. This work is supported by the  Spanish Ministry of
Science and Technology project BFM2002-02588, CIRIT project
SGR-00185, 
and the European Community projects QUPRODIS and  RESQ. E.~B.~is grateful to Brookhaven Nat.~Lab. for hospitality.

\newcommand{\PRL}[3]{Phys.~Rev.~Lett.~\textbf{#1}, #2~(#3)}
\newcommand{\PRA}[3]{Phys.~Rev. A~\textbf{#1}, #2~(#3)}
\newcommand{\JPA}[3]{J.~Phys. A~\textbf{#1}, #2~(#3)}
\newcommand{\PLA}[3]{Phys.~Lett. A~\textbf{#1}, #2~(#3)}
\newcommand{\JOB}[3]{J.~Opt. B~\textbf{#1}, #2~(#3)}
\newcommand{\JMP}[3]{J.~Math.~Phys.~\textbf{#1}, #2~(#3)}
\newcommand{\JMO}[3]{J.~Mod.~Opt.~\textbf{#1}, #2~(#3)}

 \end{document}